\documentclass[prl,nobibnotes,floatfix,superscriptaddress,twocolumn]{revtex4-2}
\usepackage{amsfonts,amsmath,graphicx,epsfig,latexsym}
\usepackage{subfigure}
\usepackage{graphicx}
\usepackage{color}
\usepackage{natbib}
\usepackage{multirow}
\usepackage{latexsym,amssymb}
\usepackage{amsmath}
\usepackage{soul}
\usepackage{ulem}
\usepackage{array}
\usepackage{siunitx}
\newcolumntype{d}[1]{D{.}{.}{#1}}
\usepackage[linktocpage,bookmarksopen,bookmarksnumbered]{hyperref}
\usepackage[dvipsnames]{xcolor}

\usepackage{comment}
\usepackage{braket}

\usepackage{multirow}

\begin{document}

\title{Self-Consistent Coulomb Interactions from Constrained Dynamical Mean-Field Theory}

\author{Antik Sihi}
\affiliation
{Department of Physics and Astronomy, West Virginia University, Morgantown, WV, USA}

\author{Subhasish Mandal}
\email{Contact author: subhasish.mandal@mail.wvu.edu}
\affiliation
{Department of Physics and Astronomy, West Virginia University, Morgantown, WV, USA}

\author{Kristjan Haule}
\affiliation{Center for Materials Theory and Department of Physics and Astronomy, Rutgers University, Piscataway, NJ 08854, United States}

%\abstract{
\begin{abstract}
{\footnotesize
We develop a self-consistent first-principles framework for determining the screened Coulomb interaction strength ($U$) based on constrained dynamical mean-field theory (cDMFT). Unlike conventional approaches, this method incorporates essential vertex corrections within the same embedded-DMFT formalism used for the electronic structure calculation.
Using the cDMFT-derived interaction strengths as input to embedded DMFT yields spectral functions in excellent agreement with photoemission experiments across a wide range of materials, spanning 3$d$ to 5$d$ transition-metal compounds, including correlated metals, Mott insulators, altermagnets, and unconventional superconductors. This unified many-body framework establishes a systematic first-principles route for determining interaction strengths in correlated materials and substantially enhances the predictive power of DFT+DMFT and its extensions.
}
\end{abstract}


\maketitle

{\it Introduction.} Strongly correlated materials containing transition metals remain a major challenge for electronic-structure theory~\cite{doi:10.1063/1.1712502,paschen2021quantum}, as their properties are often governed by correlations beyond conventional density functional theory (DFT)~\cite{onida_electronic_2002,DMFT2,DMFT3}. A common DFT extension, primarily aimed for insulating systems, is the Hubbard-corrected DFT+$U$ (or LDA+$U$) approach~\cite{Hub,Anisimov_1997,liechtenstein1995density}, which mitigates self-interaction errors and enforces piecewise linearity for localized $d$ and $f$ states~\cite{timrov2018hubbard,cococcioni2005linear,SIC_zaanen1988can,U1}. However, due to its mean-field nature, DFT+$U$ predicts infinite quasiparticle lifetimes, may spuriously open gaps by enforcing long-range order, and fails to capture dynamical correlations.

Among beyond-mean-field methods, DFT combined with dynamical mean-field theory (DFT+DMFT) has emerged as a state-of-the-art \textit{ab initio} framework for treating strong local correlations~\cite{gabi-PToday,DMFT2,DMFT3,PhysRevLett.94.026404,Kunes:2008bh,PhysRevLett.109.156402,PhysRevB.85.094505,mandal-PRL,PhysRevB.98.075155,turan-review}, while remaining computationally viable for systematic materials studies without resorting to low-energy model Hamiltonians. Despite these advantages, the predictive power of DFT+DMFT hinges on a key input parameter—the on-site screened Coulomb interaction $U$—which is often treated as adjustable parameter rather than determined self-consistently.
This limitation has so far curtailed the broader adoption of DFT+DMFT in data-driven materials discovery efforts that require reliable, high-throughput calculations and database constructions~\cite{PhysRevX.5.011006, jain2013materialsproject,curtarolo2012aflow,curtarolo2013aflowlib,JARVIS, Choudhary_ALIGNN2021, Mandal-alignn,butler2018machine,pati-ML}.

Conceptually, DFT+DMFT strikes a favorable balance by treating correlations locally while retaining statically screened interactions and essential dynamical effects. Approaches based on frequency-dependent effective interactions, such as GW+DMFT~\cite{GW_DMFT0,GW_DMFT1}, can introduce artifacts in superconducting regimes, as dynamical screening may generate spurious attractive interactions and unphysical pairing instabilities~\cite{Sham_SC0,Prokofiev_SC0}. Recent diagrammatic Monte Carlo studies of the uniform electron gas confirm that such instabilities arise with dynamical screening but are absent with static screening~\cite{Prokofiev_SC0,Prokofiev_SC1}. For unconventional superconductors, where pairing emerges from repulsive interactions and strong correlations, statically screened interactions therefore provide a more controlled and physically transparent framework, as demonstrated in recent cuprate studies~\cite{Benjamin}.

Over the past decades, several first-principles schemes have been developed to compute $U$, including constrained DFT and its linear-response extensions~\cite{Anisimov_1997,dederichs1984ground,anisimov1991density,cococcioni2005linear}, and constrained random phase approximation (cRPA)~\cite{CRPA,CRPA2,aryasetiawan2006calculations}. Alternatively, $U$ can be obtained from Hartree--Fock-based approaches~\cite{mosey2007ab,PhysRevX.5.011006}, time-dependent density functional theory~\cite{tancogne2017self}, or more recently from density functional perturbation theory-based methods~\cite{timrov2018hubbard}. However, since screening arises from charge fluctuations not fully captured within any single many-body framework, the resulting $U$ values are method- and material-dependent and generally non-transferable~\cite{PhysRevB.58.1201,tancogne2017self,tancogne2020parameter,droghetti2016fundamental}. In DFT+embedded-DMFT (eDMFT), the situation is further complicated by the mixed representation employed: local interactions and two-particle response functions are defined in a quasi-localized atomic basis, while the kinetic energy and single-particle quantities are treated in the full Kohn--Sham space. This mismatch poses a fundamental challenge for a consistent description of electronic screening and renders conventional schemes such as cDFT and cRPA conceptually incompatible with the DFT+eDMFT framework, despite their widespread use in the literature.
The accurate determination of $U$ is therefore central to establishing the first-principles character of eDMFT, particularly since other major ambiguities, such as the double-counting correction, have recently been resolved {\it exactly}~\cite{ExactDC}.

In this Letter, we present a self-consistent first-principles framework for determining the screened Coulomb interaction within the same many-body formalism used for correlated electronic structure calculations.
In the constrained DMFT (cDMFT) framework, the interaction parameter $U$ is computed self-consistently within eDMFT, ensuring internal consistency between interaction determination and the electronic structure~\cite{haule2010dynamical}.
Importantly, this formulation naturally incorporates vertex corrections arising from local dynamical correlations, which are not accounted for in conventional cDFT or cRPA schemes.

\begin{figure}[t]
\centering
\includegraphics[width=0.49\textwidth]{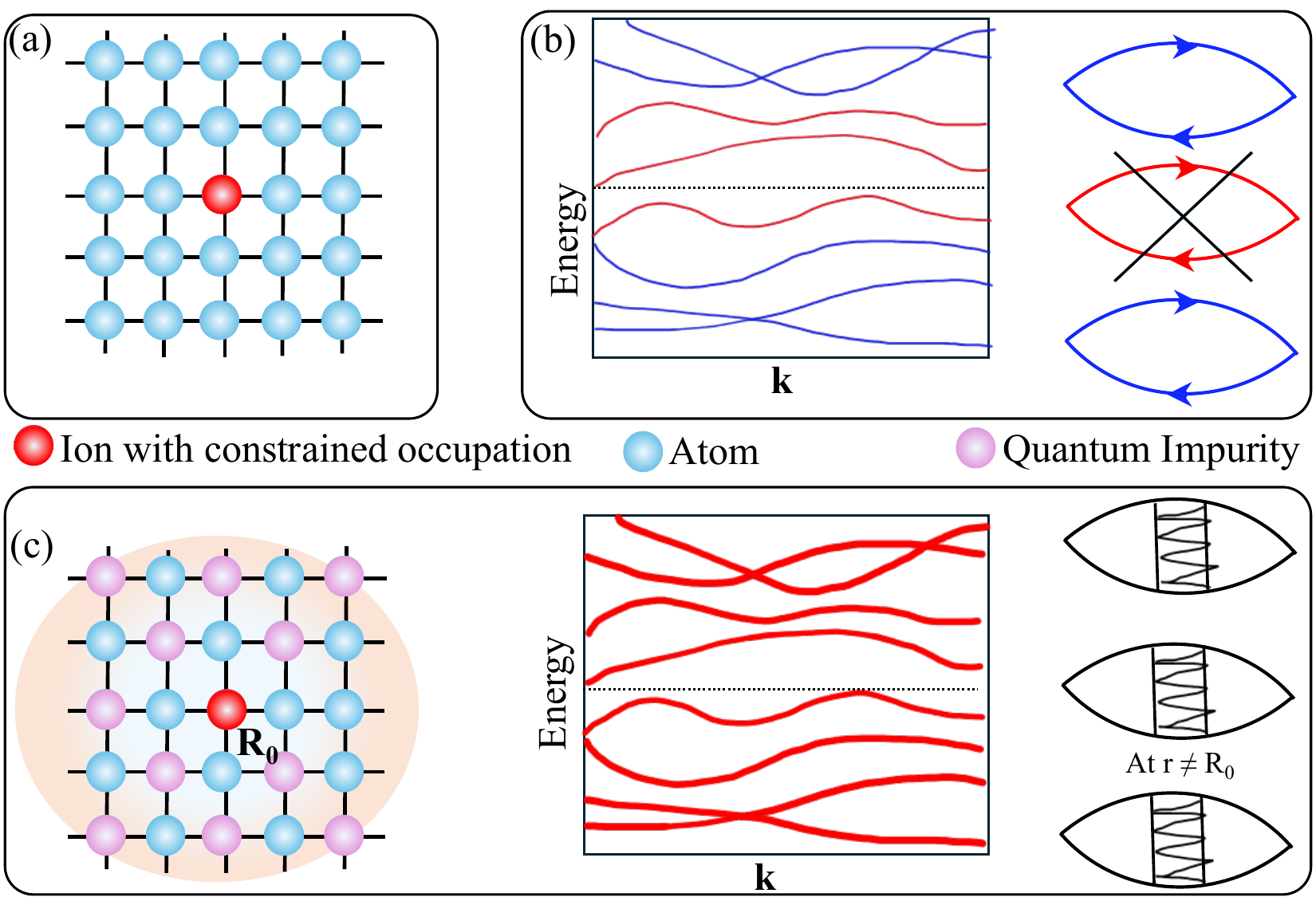}
\caption{Schematic representations of (a) cDFT, (b) cRPA, and (c) proposed cDMFT methods.}
\label{Fig1}
\end{figure}

{\it{cDMFT Method: }}To formulate the cDMFT approach, we build upon the cDFT method originally proposed by Anisimov {\it et al.}~\cite{anisimov1991density}, in which the Coulomb interaction $U$ is directly related to the local charging energy of the solid
\[
    U - \alpha J = E(N+1) - 2E(N) + E(N-1)  %= \frac{\partial^2 E}{\partial N_d^2},
\]
where $E(N+1)$, $E(N)$, and $E(N-1)$ denote the total energies corresponding to configurations with $(N+1)$, $N$, and $(N-1)$ constrained electrons in the correlated orbitals, respectively. The parameter $J$ denotes the on-site Hund’s exchange.
When rewriting the local Slater interaction employed in eDMFT functional in terms of charging-energy, spin-spin, and orbital-isospin contributions, following  Ref~\cite{Antoine_review}, the coefficient $\alpha$ can be determined explicitly. For $t_{2g}$ and $e_g$ orbitals $\alpha=1.172$ and $1.517$, respectively (details in SI \cite{Supplemental} (see also references \cite{cDMFT_tuto, cDMFT_tuto1} therein)).
 In the cDFT framework, $U$ is evaluated by expanding the crystal into a supercell [Fig.~\ref{Fig1}a] and computing total energy differences within DFT. In the present cDMFT formulation, the total energy, $E = F + T S $, is obtained from the Luttinger–Ward functional by self-consistently determining the free energy $F$ and entropy $S$ at temperature $T$ within the eDMFT framework~\cite{haule2010dynamical,haule2015free} for the supercell, using the same supercell construction. To ensure proper inclusion of screening effects within eDMFT, multiple quantum impurity problems are solved for all correlated sites in the supercell except for the central one, whose occupation is constrained. To prevent charge transfer to or from the constrained site, the hopping integrals between the constrained correlated orbitals and the surrounding sites are set to zero. This construction allows electronic screening to arise from the response of the surrounding correlated sites, which are treated on equal footing within the many-body framework. The essence of the cDFT, cRPA, and cDMFT approaches are illustrated schematically in Fig.~\ref{Fig1}(a)-(c). Panel (a) depicts the supercell-based cDFT method, panel (b) illustrates the cRPA scheme, in which screening is treated within cRPA, while vertex corrections associated with other correlated sites are neglected, and (c) shows the supercell implementation of cDMFT. In the cDMFT construction, all correlated atoms in the supercell—except for the constrained one—are treated as quantum impurities to account for local correlations. As a result, intersite screening processes are naturally included through the supercell geometry, and vertex corrections to the charge response are incorporated through the many-body evaluation of the total energy within eDMFT. In this sense, cDMFT provides a controlled extension beyond both local constrained approaches and the conventional RPA-based screening description [Fig.~\ref{Fig1}(b)].

 All calculations in this work are performed using $2\times2\times2$ supercells, except for La$_3$Ni$_2$O$_7$ and Sr$_2$IrO$_4$. Primitive unit cell of Sr$_2$IrO$_4$ is sufficiently large for this calculation and $2\times2\times1$ supercells are used for La$_3$Ni$_2$O$_7$. We have also verified the convergence with respect to supercell size by increasing the supercell to $3\times3\times3$ size and found that the $U$ changes only approximately 0.2 eV (details in SI \cite{Supplemental} (see also references \cite{eDMFT_JPSJ,ctqmc, MaxEnt, WIEN2k, sihi2020detailed} therein)). The Hund’s coupling is fixed to $J = 0.8$~eV for all systems to avoid tuning. The cDFT reference calculations are performed using the full-potential linearized augmented-plane-wave code WIEN2k~\cite{anisimov1991density}, which employs a basis consistent with what is used in eDMFT calculations (details in SI \cite{Supplemental}).

{\it{Results:}} The results obtained from cDFT and cDMFT calculations are summarized in Table~\ref{tab1} for representative materials containing 3$d$, 4$d$, and 5$d$ orbitals. We distinguish between metallic and insulating systems. For 3$d$ metals, the cDFT-calculated \( U \) lie in the range of 3–5~eV and exhibit a weak decrease when moving from 3$d$ to 5$d$ systems, reflecting the increasing spatial extent of the correlated orbitals. For the same materials, cDMFT systematically yields larger interaction strengths, with $U \approx 5$–$6$~eV for metals and $7$–$10$~eV for insulators.
For 4$d$ and 5$d$ compounds, cDFT predicts relatively small and nearly material-independent values, $U \sim 3$~eV, for both metals and insulators.
In contrast, cDMFT yields a clear separation between metals and insulators, with $U \sim 6$~eV and $\sim5$~eV for 4$d$ and 5$d$ insulators, respectively, and $U \sim 4$~eV for the corresponding metals. This separation reflects the different degree of correlation-induced localization captured self-consistently within the eDMFT framework.

\begin{table}[]
\caption{Calculated values of $U$ (in eV) using cDFT and cDMFT for various class of systems and the Hund's coupling ($J$) is fixed at 0.8 eV for all the calculations.}\label{tab1}
\begin{tabular*}{0.49\textwidth}{@{\extracolsep\fill}lccccc}
\toprule%
 Systems & Materials & \multicolumn{2}{c}{Computed values of $U$} \\[1mm]
  &  & cDFT (eV) & cDMFT (eV)\\[1mm]
%\midrule
\hline
%\midrule
& La$_3$Ni$_2$O$_7$ (LP) & 5.3 & 6.3 \\[1mm]
& La$_3$Ni$_2$O$_7$ (HP) & 4.6 & 5.5 \\[1mm]
3$d$ metal & V$_2$O$_3$ & 5.3 & 6.3 \\[1mm]
& FeSe & 4.9 & 5.7 \\[1mm]
& Vanadium (V) & 3.7 & 3.8 \\[1mm]
\hline
& MnTe & 4.4 & 6.7 \\[1mm]
3$d$ insulator & V$_2$O$_3$ & 5.4 & 8.5 \\[1mm]
& NiO & 5.9 & 8.5 \\[1mm]
& CoO & 4.5 & 10.2 \\[1mm]
\hline
4$d$ metal & RuO$_2$ & 3.0 & 4.0 \\[1mm]
\hline
5$d$ metal & SrIrO$_3$ & 3.3 & 3.8 \\[1mm]
5$d$ insulator & Sr$_2$IrO$_4$ & 3.2 & 5.3 \\[1mm]
\botrule
\end{tabular*}
\end{table}

\begin{figure*}[t]
\centering
\includegraphics[width=1.0\textwidth]{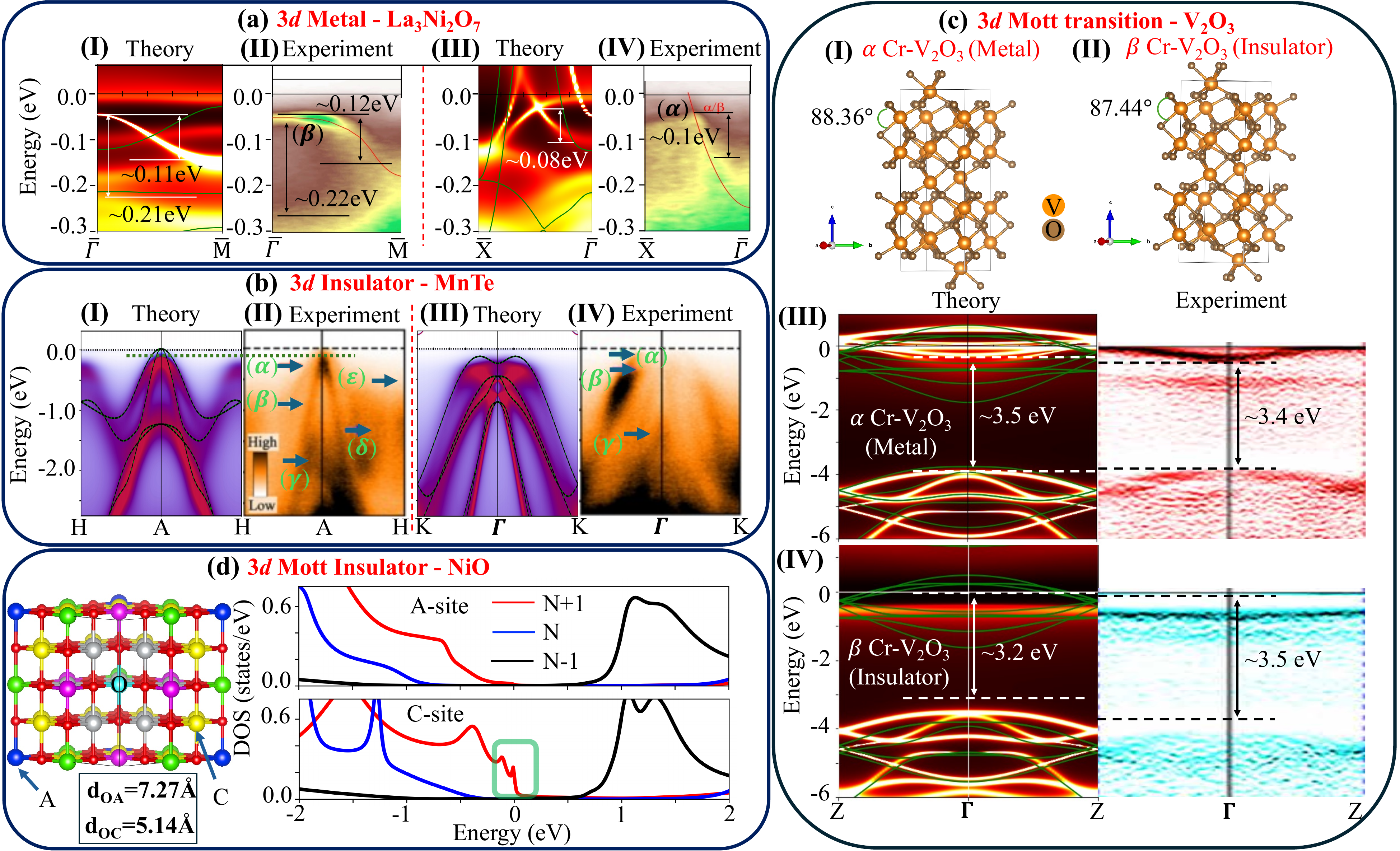}
\caption{ Application of cDMFT method for various 3$d$ metals and insulators. (a) Spectral functions comparison of La$_3$Ni$_2$O$_7$ between eDMFT (I \& III) and experiment (II \& IV)~\cite{LNO_exp_yang2024}. The bandwidths are marked by double sided arrow. Thin green lines denote the band structures obtained from DFT+$U$ calculations. (b) eDMFT computed spin and momentum resolved spectral functions of MnTe (I \& III) compared with the experimental ARPES spectra (II and IV) \cite{PRB_MnTe_exp}. In computed spectra, the minority (majority) electrons are denoted by red (blue) color. The green solid (black dotted) lines represent the spin-up (spin-down) bands from DFT+$U$ computation. (c) Crystal structures (I and II) and spectral functions comparisons (III and IV) of V$_2$O$_3$ in $\alpha$ and $\beta$ phases to show the metal-insulator transition. Experimental spectra are taken from Ref.~\cite{v2o3_exp_MIT_sciadv}. DFT+$U$ bands are marked with green lines. (d) Site dependent projected density of states (PDOS) of Ni-3$d$ orbitals in a 2$\times$2$\times$2 supercells for N+1, N and N-1 constrained electrons are plotted for two different distances (\lq OA\rq and \lq OC\rq) from origin (\lq O\rq), showing appearance of the Kondo peak near E$_F$, which is marked by a green box.}\label{fig_3d}
\end{figure*}

{\it 3$d$ metals:} We examine several 3$d$ transition-metal compounds (Fig.~\ref{fig_3d}) by comparing experimental angle-resolved photoemission spectrosocy (ARPES) data with eDMFT spectral functions (SFs) (thick lines) and DFT+$U$ (thin lines) band structures. The interaction parameters \(U\) are obtained from cDMFT and cDFT, respectively. As a representative example, we focus on La$_3$Ni$_2$O$_7$ (LNO), a recently discovered high-\(T_c\) superconductor that remains metallic in both low- and high-pressure phases. The calculated \(U\) values show moderate pressure dependence in cDMFT and cDFT. ARPES spectra at ambient pressure~\cite{LNO_exp_yang2024} are compared with eDMFT and DFT+$U$ results along $\bar{\Gamma}$–$\bar{M}$ and $\bar{X}$–$\bar{\Gamma}$ (Fig.~\ref{fig_3d}(a)). To account for uncertainty in the experimental doping, the Fermi level (E$_F$) was shifted by 0.035~eV. DFT+$U$ significantly overestimates the bandwidth and fails to capture the observed mass renormalization, whereas eDMFT reproduces it accurately. For example, the $\alpha$-band bandwidth along $\bar{\Gamma}$–$\bar{X}$ is $\sim$0.1~eV experimentally and 0.08~eV in eDMFT, while the $\beta$-band along $\bar{M}$–$\bar{\Gamma}$–$\bar{M}$ shows excellent agreement (0.12~eV vs.\ 0.11~eV). The energy separation of the two band maxima at $\bar{\Gamma}$ also agrees well (0.22~eV experiment vs.\ 0.21~eV eDMFT). The role of pressure‑dependent correlations in LNO has also been investigated recently using eDMFT with a similar value of $U$ \cite{bleys2025role}.

Beyond LNO, eDMFT has been successfully applied to other 3$d$ metals, including Fe-based pnictides and chalcogenides such as monolayer FeSe and elemental Fe, using \(U \sim 5\)~eV. In these systems, eDMFT achieves excellent agreement with experiments for spectral properties, lattice dynamics, and electron--phonon coupling~\cite{mandal2014strong,mandal-PRL,yin2011kinetic,JPSJ,phonon-KH,Antik-EPC}.

{\it 3$d$ insulators:} We consider the 3$d$ altermagnetic insulator MnTe~\cite{PRB_MnTe_exp}, which breaks time-reversal symmetry and exhibits spin-split bands without a net magnetic moment. Spin- and momentum-resolved SFs from eDMFT, together with spin-resolved DFT+$U$ bands, are shown in Figs.~\ref{fig_3d}(b)(I),(III) and compared with ARPES data in Figs.~\ref{fig_3d}(b)(II),(IV). While the spectra are largely spin-degenerate along $H$–$A$–$H$ and $K$–$\Gamma$–$K$ paths, eDMFT correctly captures the lifting of spin degeneracy along $\Gamma$–$L$~\cite{wan2025high}, consistent with experiment~\cite{PRB_MnTe_exp}. Near the $A$ point, two hole-like bands ($\alpha$, $\beta$) observed in ARPES are well reproduced by eDMFT. A dome-shaped feature $\gamma$ centered near $-1.8$~eV is also captured, with band maxima at $-1.4$~eV in eDMFT and $-1.25$~eV in DFT+$U$. Additional broader features ($\delta$, $\epsilon$) appear in eDMFT, though some discrepancies remain near the $H$ point. Along $K$–$\Gamma$–$K$, an inverted $V$-shaped band ($\alpha$) and two inner bands ($\beta$, $\gamma$) with maxima near $-0.5$ and $-1.4$~eV agree reasonably with ARPES. Overall, eDMFT captures the dominant experimental features and provides improved bandwidth renormalization compared to DFT+$U$.

Other 3$d$ Mott insulators, including the binary transition-metal oxides NiO, CoO, FeO, and MnO, have been systematically analyzed in Ref.~\cite{mandal2019systematic}, where eDMFT calculations with $U=10$~eV showed excellent agreement with both photoemission and optical absorption measurements.

{\it{3$d$-MIT: }}Next, we examine the metal–insulator transition (MIT) in V$_2$O$_3$, a prototypical Mott–Hubbard system in which electronic correlations and lattice degrees of freedom play comparably important roles. In its corundum structure (space group R$\Bar{3}$c), V$_2$O$_3$ is a paramagnetic metal (PM) under ambient conditions and transforms into a paramagnetic insulator (PI) upon light Cr doping, as in (V$_{0.962}$Cr$_{0.038}$)$_2$O$_3$. Here, we investigate the corresponding iso-structural MIT within the corundum phase.

The ARPES spectra, eDMFT SFs, and DFT+$U$ band structures are summarized in Fig.~\ref{fig_3d}(c). The metallic and insulating phases are denoted as $\alpha$-V$_2$O$_3$ and $\beta$-V$_2$O$_3$, respectively, for which the cDMFT-derived interaction $U$ are 6.3 and 8.5 eV, respectively (Table~\ref{tab1}). This pronounced phase dependence of $U$ highlights the importance of self-consistent screening feedback across the MIT, which goes beyond a fixed-$U$ Hubbard description and is essential for achieving quantitative agreement with ARPES.
The cDMFT values of $U$ are significantly larger than those obtained by cRPA, where $U$ ranges from $2$ to $4.9$~eV depending on the choice of projected orbitals used in the calculation~\cite{PhysRevB.110.045117}.

To elucidate the mechanism of MIT, we optimize the atomic positions of both phases within eDMFT, which includes the electronic entropy contributions in the force evaluation, and is a essential to reproduce MIT within eDMFT.
The resulting structures, shown schematically in Figs.~\ref{fig_3d}(c)(I,II), exhibit tetrahedral angles of 88.4\textdegree{} and 87.4\textdegree{} for the metallic ($\alpha$) and insulating ($\beta$) phases, respectively (see SI \cite{Supplemental} for details), which agree extremely well with the corresponding experimental angle  of 88.36\textdegree{} (87.47\textdegree{}) for the metallic (insulating) phase \cite{Robinson1975}.

The corresponding eDMFT SFs are shown in Figs.~\ref{fig_3d}(c)(III,IV) alongside ARPES data~\cite{v2o3_exp_MIT_sciadv}.
Unlike DFT+$U$, which fails to reproduce the MIT without imposing antiferromagnetic order, eDMFT captures the transition naturally within the paramagnetic phase.
In the metallic $\alpha$-phase, both ARPES and eDMFT show a dispersive quasiparticle band crossing the E$_F$ along $\Gamma$–Z, consistent with metallic behavior, as well as a deeper valence band at approximately $-3.5$~eV dominated by O-$p$ states. The MIT  is also reflected in the evolution of the electronic self-energy (see SI \cite{Supplemental}). Overall, the spectral features and bandwidths show excellent agreement between eDMFT and experiment.
In the insulating $\beta$-phase, ARPES reveals a nondispersive quasi-localized state near $-0.7$~eV, which is accurately reproduced by eDMFT, demonstrating its ability to describe both itinerant and localized electronic states across the MIT.

\textit{cDMFT in Mott state: }To gain microscopic insight into why cDMFT yields substantially larger interaction strengths in Mott insulating states than cDFT or cRPA, we consider the prototypical charge-transfer Mott insulator NiO and examine its electronic structure upon adding or removing a charge at the central constrained site, as implemented in cDMFT.
Fig.~\ref{fig_3d}(d) shows the site-resolved projected density of states (PDOS) of Ni $3d$ orbitals for ions at sites A and C, located at distances $d_{OA}$ and $d_{OC}$ from the central constrained Ni site (denoted O) within a $2\times2\times2$ supercell.
The PDOS are computed for configurations constrained to $(N+1)$, $N$, and $(N-1)$ electrons.
In the nominal ($N$) configuration, a clear insulating gap is observed, consistent with the Mott-insulating ground state.
Upon adding an electron to the constrained site ($(N+1)$ configuration), a sharp resonance develops at the E$_F$ on the nearest-neighbor sites (C), signaling the emergence of a Kondo-like quasiparticle peak associated with the screening of localized Ni $3d$ moments.
This screening cloud is highly nonlocal and spatially inhomogeneous: the quasiparticle resonance rapidly diminishes on more distant sites (A), where the insulating gap is restored.

This behavior reveals the microscopic origin of the enhanced screening captured by cDMFT.
Charge fluctuations on the constrained site induce a quasi-metallic screening environment on nearby correlated sites, a genuinely many-body effect that cannot be described within static constrained approaches.
As a consequence, the effective $U$ in cDMFT is significantly larger than that obtained from cDFT (see SI \cite{Supplemental}) or cRPA, irrespective of the value of $U$ employed in DFT+$U$ calculations.
Such spatially resolved, correlation-driven screening processes have not been accessible within conventional constrained approaches and are naturally revealed within the cDMFT framework.

\begin{figure}[]
\centering
\includegraphics[width=0.49\textwidth]{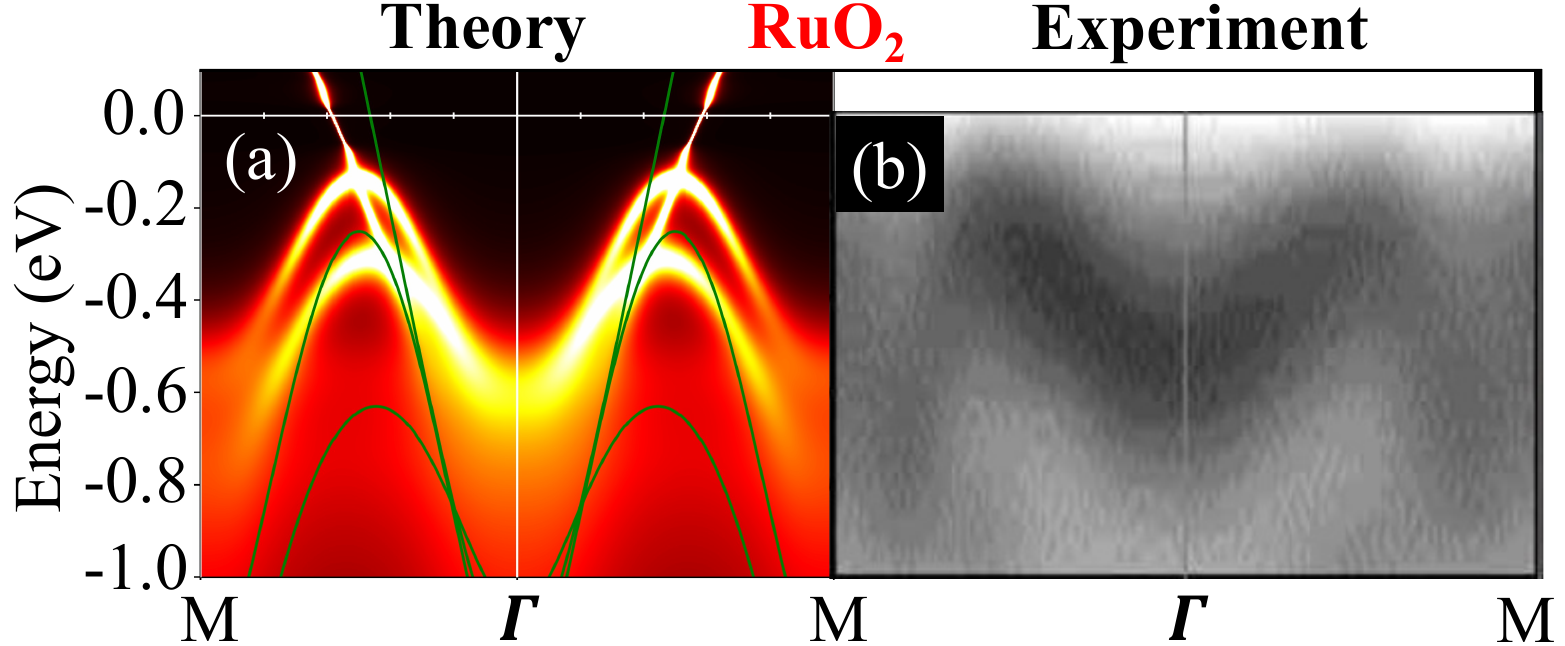}
\caption{Comparison of spectral functions obtained from (a) eDMFT in the paramagnetic phase and (b) ARPES measurements of 4$d$-metallic RuO$_2$ (reproduced from Ref.~\cite{ruo2_exp_fedchenko2024}). The green lines in (a) denote the dispersion obtained from DFT+$U$ calculations.
} \label{fig_ruo2}
\end{figure}

{\it{Spectral function comparison for 4$d$-compounds: }} We compared the eDMFT spectral functions with experiments for 4$d$  materials, as shown in Figs.~\ref{fig_ruo2}(a,b). Here we consider RuO$_2$ in its paramagnetic metallic phase \cite{kessler2024absence}. The ARPES data, taken from Ref~\cite{ruo2_exp_fedchenko2024}, show overall good agreement with the eDMFT spectral functions. While the ARPES bands are quite broad, the `M'-shaped feature, the bandwidths are clearly in agreement in eDMFT. In contrast, DFT+$U$ yields a significantly larger dispersion and substantially overestimates the bandwidth(in thin green lines).

\begin{figure}[]
\centering
\includegraphics[width=0.49\textwidth]{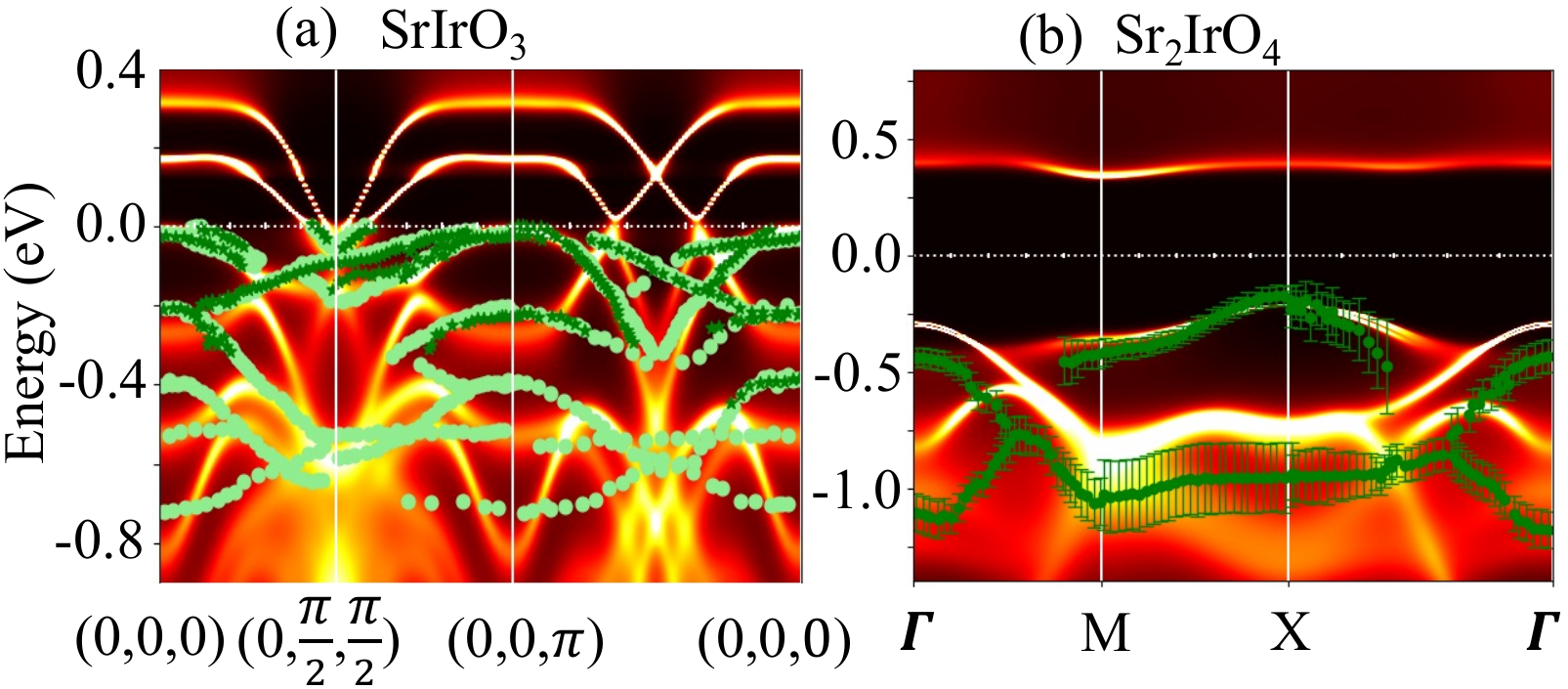}
\caption{Momentum resolved spectral functions from eDMFT for (a) SrIrO$_3$ and (b) Sr$_2$IrO$_4$. ARPES data (green) and the second derivative data (lime green) of (a) SrIrO$_3$ and (b) Sr$_2$IrO$_4$ are reproduced from \cite{nie2015interplay} for comparison.}\label{fig_5d_electrons}
\end{figure}

{\it{Spectral function comparison for 5$d$-compounds: }}We also examine representative 5$d$ systems, selecting the Ir-based compounds SrIrO$_3$ and Sr$_2$IrO$_4$ as prototypical examples of a correlated metal and insulator, respectively.
Figs.~\ref{fig_5d_electrons}(a,b) compare the eDMFT spectral functions for metallic SrIrO$_3$ and insulating Sr$_2$IrO$_4$ with the corresponding experimental ARPES spectra from Ref.~\cite{nie2015interplay}. In both the cases, eDMFT spectra match very well with low energy experimental spectra, while there is slight mismatch for the higher energy bands.

Thus, together, self-consistent cDMFT and eDMFT form a robust framework that enables more accurate, tuning-free, and consistent agreement with experiments across a broad class of correlated systems.

%\begin{center}

\textit{Sensitivity and transferability of $U$: }In Supplemental Fig.~S3~\cite{Supplemental}, we illustrate the dependence on $U$ in DFT+$U$ and eDMFT calculations for NiO.
In a recent study, we further examined the $U$ dependence for the semiconducting altermagnet MnTe and the metallic CrSb~\cite{wan2025high}.
Overall, the eDMFT SFs exhibit a noticeably reduced sensitivity to moderate variations of $U$ compared to DFT+$U$.
This reduced sensitivity should not be interpreted as a diminished role of the interaction strength, but rather as a consequence of the internally consistent treatment of correlations and screening within eDMFT.
Once $U$ is determined on a first-principles footing, small deviations around its optimal value do not lead to qualitative changes in the low-energy spectra.
For example, a range of metallic systems are well described using $U=5$~eV~\cite{Metal_paper}, while photoemission and ARPES spectra for several binary transition-metal oxides are accurately reproduced with $U=10$~eV, even though the cDMFT-computed interaction strengths span a range from $8.5$ to approximately $10.5$~eV~\cite{TMO2-SM}.
This robustness allows a single interaction strength to be employed across closely related families of materials, such as ABO$_3$ perovskite oxides (A = Ca, Sr, La; B = V, Cr, Mn, Fe, Co, Ni), without compromising predictive accuracy.
By contrast, in cDFT and cRPA the computed values of $U$ often exhibit a pronounced and nonlinear dependence on $d$-electron occupancy at the transition-metal site~\cite{crpa-abo3}, which complicates their use in systematic or high-throughput studies.

\textit{Conclusion:} We have introduced a self-consistent framework for determining the screened Coulomb interaction within constrained dynamical mean-field theory (cDMFT), ensuring internal consistency between interaction parameters and the correlated electronic structure. By evaluating the interaction strength within the same many-body formalism used in eDMFT, the approach naturally incorporates local vertex corrections, which are absent in conventional constrained schemes such as cDFT and cRPA.
Across a broad class of 3d–5d transition-metal systems, including correlated metals and insulators, oxides, altermagnets, and unconventional superconductors, the resulting interaction strengths lead to spectral functions in close agreement with photoemission and ARPES experiments. Despite the locality of the vertex corrections, we find that they capture the dominant screening processes relevant for low-energy correlated physics, yielding interaction strengths that are robust and transferable within material families.
More generally, these results suggest that local vertex corrections, when treated self-consistently within a DMFT framework, are sufficient to capture the essential screening physics required for accurate effective interactions. This insight is not limited to eDMFT, but is expected to be relevant for a broad class of approaches that build on DMFT and their extensions. These results establish cDMFT as a practical and internally consistent route for determining effective Coulomb interactions in correlated materials, significantly enhancing the predictive capability of DMFT-based electronic structure methods without introducing additional empirical parameters.

\section{Acknowledgments}

Authors acknowledge the support from the National Science Foundation Award No. NSF OAC-2311557 and NSF OAC-2311558. Authors benefited from the Frontera supercomputer at the Texas Advanced Computing Center (TACC) at The University of Texas at Austin, which is supported by National Science Foundation Grant No. OAC-1818253.
KH also acknowledges support of NSF DMR-2233892.

\section{Data availability}
All data that support the plots within this paper and other findings of this study are available from the corresponding author upon reasonable request. Source data are provided with this manuscript.

%\bibliographystyle{apsrev4-2} % PRL uses apsrev4-2
%\bibliography{ref} % common bib file

\clearpage
\onecolumngrid

\section*{Supplementary Information: Self-Consistent Coulomb Interactions from Constrained Dynamical Mean-Field Theory}

\section{Theoretical Descriptions of constrained dynamical mean-field theory}
The form of the Slater parametrization of the Coulomb $U$ is given below \cite{georges2013strong,cDMFT_tuto,cDMFT_tuto1}:
\begin{align*}
\begin{split}
    \hat{U} = \frac{1}{2}\sum_{\{m \},s,s'} \Biggl \langle Y_{lm_1}(r)\frac{u_l(r)}{r}Y_{lm_2}(r')\frac{u_l(r')}{r'} | V_{DMFT} (r - r') | Y_{lm_3}(r')\frac{u_l(r')}{r'}Y_{lm_4}(r)\frac{u_l(r)}{r}  \biggr \rangle \psi^\dagger_{m_1s}\psi^\dagger_{m_2s'}\psi_{m_3s'}\psi_{m_4s} \\
    = \frac{1}{2}\sum_{\{m \},s,s'} U_{m_1m_2m_3m_4}\psi^\dagger_{m_1s}\psi^\dagger_{m_2s'}\psi_{m_3s'}\psi_{m_4s}
\end{split}
\end{align*}

where $u_l(r)$ is the radial function and $Y_{lm}(r)$ represents the spherical harmonics. The form of $V_{DMFT}(r-r')$ with considering the Yukawa screening is provided below:

\begin{equation*}
    V_{DMFT}(r-r') = \frac{e^{-\lambda|r-r'|}}{|r-r'|}
\end{equation*}

and the Coulomb repulsion matrix elements $U_{m_1m_2m_3m_4}$ takes the form:

\begin{equation*}
    U_{m_1m_2m_3m_4} = \sum_{m,k}\frac{4\pi}{2k+1} \langle Y_{lm_1}| Y_{km} | Y_{lm_4} \rangle \langle Y_{lm_2}| Y_{km}^* | Y_{lm_3} \rangle F^k
\end{equation*}

where $F^k$ denotes the Slater parameters. In case of entire $d$-shell, it is well known that the relation in between the Coulomb interaction $U_S$ = $F^0$ and Hund's coupling $J_S$ = $\frac{F^2+F^4}{14}$ is found to be in the Coulomb part of the Hamiltonian as,

\begin{equation*}
    H = (U_S - \alpha J_S) \frac{N(N-1)}{2} - \alpha 'J_S\vec{S^2} + \dots
\end{equation*}

where $N$ and $S$ are the total charge and spin, respectively. The values of $\alpha$ parameter for $t_{2g}$ ($e_g$) subshell and total $d$ orbitals are 1.172 (1.517) and 1.15, respectively \cite{cDMFT_tuto}. Only the first term in the above equation is important for paramagnetic calculation within eDMFT. Now in order to calculate the first term of the above equation using constrained dynamical mean-field theory (cDMFT), we evaluate the Coulomb interaction Hubbard $U$ using the following expression \cite{cDMFT_tuto}:

\begin{equation*}
    U_S - \alpha J_S = E[N+1] + E[N-1] - 2E[N]
\end{equation*}

where, $E(N+1)$, $E(N)$, and $E(N-1)$ are representing the total energies corresponding to $(N+1)$, $N$, and $(N-1)$ constrained electrons in the correlated orbitals (for the $d$ or $f$ electrons). We used the energy difference rather than the derivatives because we don't have any Janak's theorem in cDMFT as compare to cDFT method \cite{cDMFT_tuto}.

\section{Computational Details}

Here, the eDMFT code is used for all density functional theory (DFT) + embedded dynamical mean-field theory (DFT+eDMFT) computations and cDMFT calculations \cite{eDMFT_JPSJ,eDMFT2010}. WIEN2k, which is an all-electron full-potential linearized augmented plane-wave (FP-LAPW) code, is used to perform the cDFT and DFT part of the DFT+eDMFT calculations \cite{WIEN2k}. In order to carry out the cDMFT calculations, three configurations of N+1, N and N-1 electrons are prepared by keeping the conduction electrons of $d$ or $f$ of one central transition metal ion (3$d$, 4$d$ and 5$d$) in core states of the corresponding supercell structure as similar to cDFT proposed by Anisimov $et$ $al$ \cite{anisimov1991density}. Other remaining transition ions of the supercell are considered as separate quantum impurity problems and solved it using the continuous-time quantum Monte Carlo (CTQMC) method \cite{ctqmc}. Here, it is crucial to note that the total energies of each configurations for computing $U$ within cDMFT are estimated by the charge self-consistent eDMFT calculations. \textit{exact} double counting (DC) and density-density type Coulomb interaction used for all the cDMFT calculation along with $\beta$=50 eV$^{-1}$ (T=230 K) \cite{ExactDC}. Typically, it is found that \textit{nominal} DC gives 3-4\% higher values of computed $U$ than \textit{exact}. After obtained the $U$ parameter, we compute standard DFT+eDMFT calculation for conventional unit cell of each materials with full rotionally invariant Coulomb interaction and compare with available experimental photoemission data. To obtain the eDMFT computed spectral functions on real axis, analytical continuation is performed using maximum entropy method~\cite{MaxEnt}. The detailed information of crystal structures and the number of impurity solvers of all the studied materials are provided in Supplementary Table \ref{tab2}. The optimized atomic positions of $\alpha$-V$_2$O$_3$ and $\beta$-V$_2$O$_3$ using eDMFT with cDMFT calculated $U$ are given in Supplementary Table \ref{tab3}

%Multiples impurity solvers are considered for the cDMFT calculation

\begin{table}[]
\caption{Detailed crystal information of materials and the number of impurity solvers used in cDMFT calculations for the supercell structure.}\label{tab2}
\begin{tabular*}{0.9\textwidth}{@{\extracolsep\fill}lccccccccc}
\toprule%
 Systems & Materials & Space-group & &Lattice parameters & & & Number of impurity solvers \\[1mm]
%\midrule
\hline \\[0.5mm]
%\midrule
&  &  & & a=5.4392 \AA &  &  &  & \\[1mm]
& La$_3$Ni$_2$O$_7$ (LP) & Amam & & b=5.3768 \AA &  &  & 7 \\[1mm]
&  &  & & c=20.403 \AA &  &  &  & \\[2mm]

&  &  & & a=5.289 \AA &  &  &  & \\[1mm]
3$d$ metal & La$_3$Ni$_2$O$_7$ (HP) & Fmmm & & b=5.218 \AA &  &  & 7 \\[1mm]
&  &  & & c=19.734 \AA &  &  &  & \\[2mm]

&  &  & & a=4.954 \AA &  &  &  & \\[1mm]
& V$_2$O$_3$ & R-3c & & b=4.954 \AA &  &  & 11 \\[1mm]
&  &  & & c=13.9906 \AA &  &  &  & \\[2mm]

&  &  & & a=3.7724 \AA &  &  &  & \\[1mm]
& FeSe & P4/nmm & & b=3.7724 \AA & &  & 7 \\[1mm]
&  &  & & c=5.5217 \AA &  &  &  & \\[2mm]

\hline \\[0.5mm]

&  &  & & a=4.1475 \AA &  &  &  & \\[1mm]
& MnTe & P6$_3$/mmc & & b=4.1475 \AA & &  & 5 \\[1mm]
&  &  & & c=6.710 \AA &  &  &  & \\[2mm]

&  &  & & a=5.0018 \AA &  &  &  & \\[1mm]
3$d$ insulator & V$_2$O$_3$ & R-3c & & b=5.0018 \AA &  &  & 11 \\[1mm]
&  &  & & c=13.9238 \AA &  &  &  & \\[2mm]

&  &  & & a=4.1943 \AA &  &  &  & \\[1mm]
& NiO & Fm-3m & & b=4.1943 \AA & &  & 5 \\[1mm]
&  &  & & c=4.1943 \AA &  &  &  & \\[2mm]

&  &  & & a=4.254 \AA &  &  &  & \\[1mm]
& CoO & Fm-3m & & b=4.254 \AA & &  & 5 \\[1mm]
&  &  & & c=4.254 \AA &  &  &  & \\[2mm]

\hline \\[0.5mm]

&  &  & & a=4.492 \AA &  &  &  & \\[1mm]
4$d$ metal & RuO$_2$ & P4$_2$/mnm & & b=4.492 \AA & &  & 7 \\[1mm]
&  &  & & c=3.1061 \AA &  &  &  & \\[2mm]

\hline \\[0.5mm]

&  &  & & a=5.5923 \AA &  &  &  & \\[1mm]
5$d$ metal & SrIrO$_3$ & Pnma & & b=7.884 \AA & &  & 11 \\[1mm]
&  &  & & c=5.563 \AA &  &  &  & \\[2mm]

&  &  & & a=5.5184 \AA &  &  &  & \\[1mm]
5$d$ insulator & Sr$_2$IrO$_4$ & Pcca & & b=5.5184 \AA & &  & 5 \\[1mm]
&  &  & & c=26.2235 \AA &  &  &  & \\[2mm]

\botrule
\end{tabular*}
\end{table}

\begin{table}[]
\caption{Optimized atomic positions of $\alpha$-V$_2$O$_3$ and $\beta$-V$_2$O$_3$ using eDMFT, where V is sitting at 4c site [(x,x,x),(-x+1/2,-x+1/2,-x+1/2),(-x,-x,-x),(x+1/2,x+1/2,x+1/2) and O is situated at 6e site [(y,-y+ 1/2,1/4),(1/4,y,-y+1/2),(-y+1/2,1/4,y),(-y,y+1/2,3/4),(3/4,-y,y+1/2),(y+1/2,3/4,-y)]. }\label{tab3}
\begin{tabular*}{0.49\textwidth}{@{\extracolsep\fill}lccccccccc}
\toprule%
& & Materials & & V-site & & O-site & & & \\ [1mm]
\hline
&  & $\alpha$-V$_2$O$_3$ (Metal) & & x=0.34648000 &  & y=0.56139999 & \\[1mm]
& & $\beta$-V$_2$O$_3$ (Insulator) & & x=0.34849182 &  & y=0.55640218 & \\[1mm]
\botrule
\end{tabular*}
\end{table}

\clearpage

\begin{figure*}[t]
\centering
\includegraphics[width=0.99\textwidth]{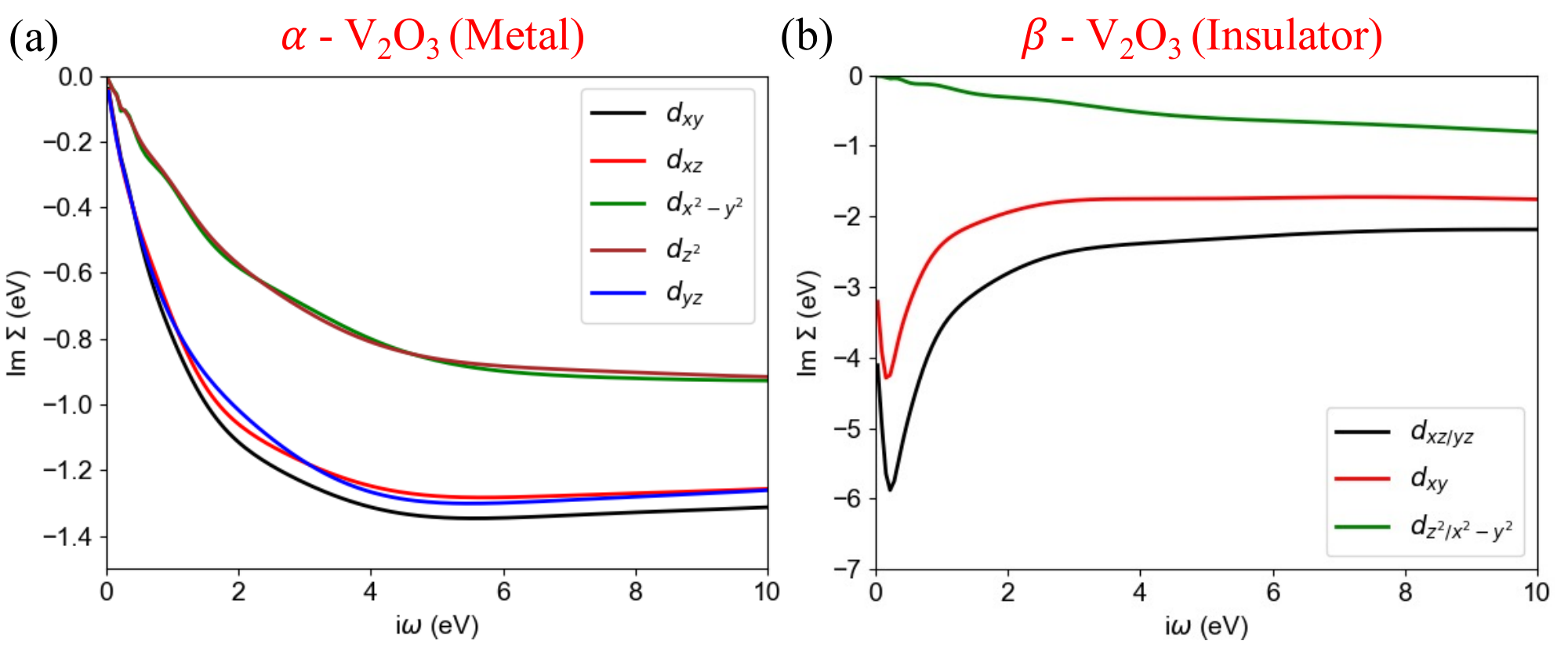}
\caption{Orbital-resolved imaginary part of self-energy in Matsubara frequency ($i\omega$) for (a) $\alpha$-V$_2$O$_3$ in metallic phase and (b) $\beta$-V$_2$O$_3$ in insulating phase.}\label{v2o3_selfenergy}
\end{figure*}

\section{Self-energy description of V$_2$O$_3$}

The metal to insulator transition (MIT) of V$_2$O$_3$ can be examined directly from the orbital resolved imaginary part of self-energy ($Im\Sigma$) as a function of imaginary frequency ($i\omega$), which is directly accessible from CTQMC. The corresponding figures of $Im\Sigma$($i\omega$) for metallic and insulating phases of V$_2$O$_3$ are plotted in Figs. S1(a) and (b), respectively. In Fig. \ref{v2o3_selfenergy}(a), it is observed that the values of $Im\Sigma$($i\omega$) for all $d$-orbitals are approaching to zero at $i\omega$=0.0. However, in case of Fig. \ref{v2o3_selfenergy}(b), the values of $Im\Sigma$($i\omega$) at $i\omega$=0.0 of $d_{xz/yz}$ and $d_{xy}$ orbitals are found to be $\sim$-3.1 eV and $\sim$-4 eV. The later one $\beta$-V$_2$O$_3$ shows insulating phase because of non-zero values of $Im\Sigma$($i\omega$) at $i\omega$=0.0, where due to $Im\Sigma$($i\omega$)$\to$0  at $i\omega$=0.0, the metallic spectral function is obtained for $\alpha$-V$_2$O$_3$ as described in main text.

\begin{figure*}[t]
\centering
\includegraphics[width=0.99\textwidth]{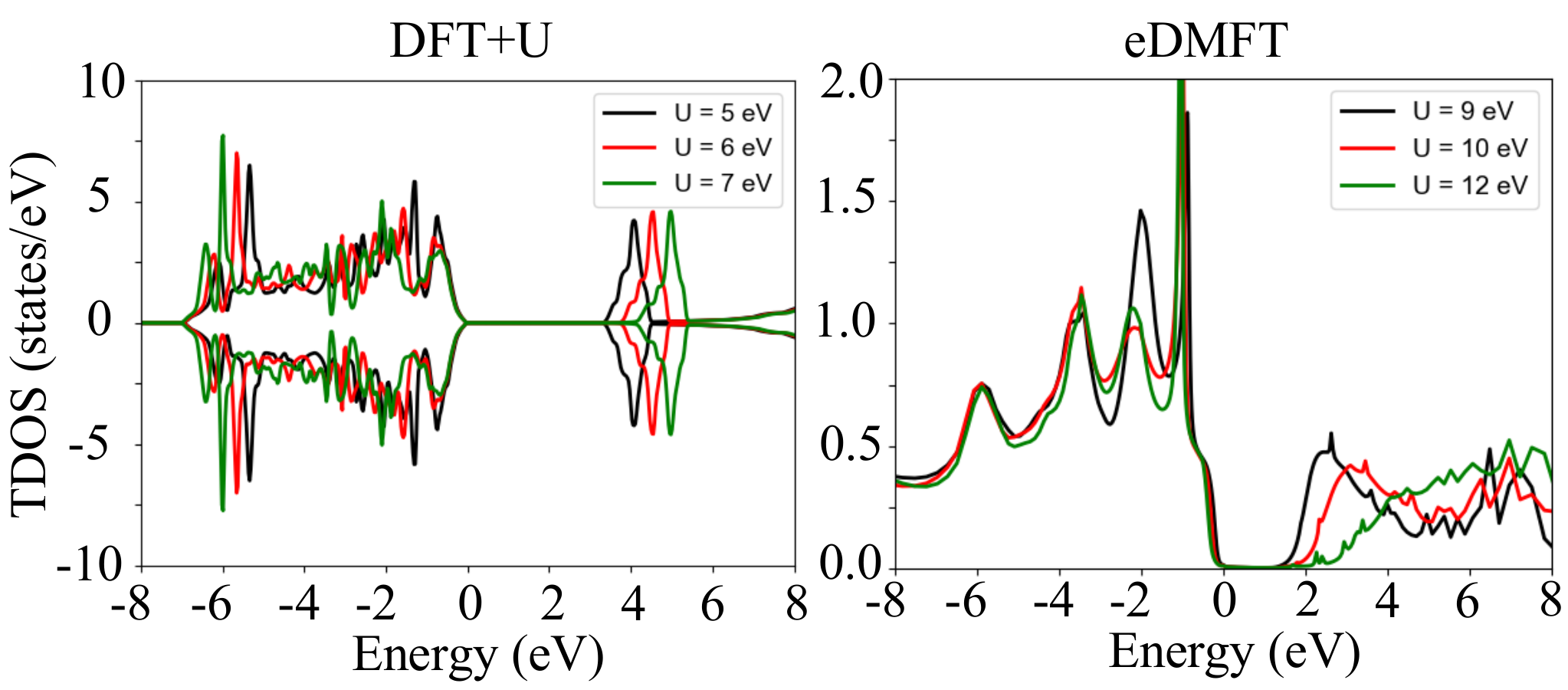}
\caption{Total density of states (TDOS) of NiO by (a) DFT+$U$ calculations with parameter $U$=5.0 eV (black), $U$=6.0 eV (red), $U$=7.0 eV (green) and (b) eDMFT calculations using $U$=9.0 eV (black), $U$=10.0 eV (red), $U$=12.0 eV (green).  }\label{fig_U_sensitivity}
\end{figure*}

\section{Sensitivity of $U$ in eDMFT compared to DFT+$U$}

In order to examine the sensitivity of $U$ in electronic structure, the direct comparison of total density of states (TDOS) obtained from DFT+$U$ and eDMFT calculations for NiO are shown in Figs. S2(a) and (b), respectively. Anti-ferromagnetic (AFM) structure of NiO is considered for DFT+$U$ calculation, whereas paramagnetic calculations performed for eDMFT as the gap size does not change across the Mott transition as shown in the earlier work for NiO \cite{TMO2-SM}. We find that eDMFT gap size is quite insensitive to the $U$-value as we change it from 9 to 12 eV. However, in DFT+$U$, as expected the gap size changes 0.4 eV going from 5 to 7 eV. Hence given 1 eV of change in $U$, the change in the gap size is 1.6 times more in DFT as compared to eDMFT for NiO.
%calculations with $U$=5-8 eV, the major peak positions and band gaps of TDOS are changing much.

\begin{figure*}[t]
\centering
\includegraphics[width=0.99\textwidth]{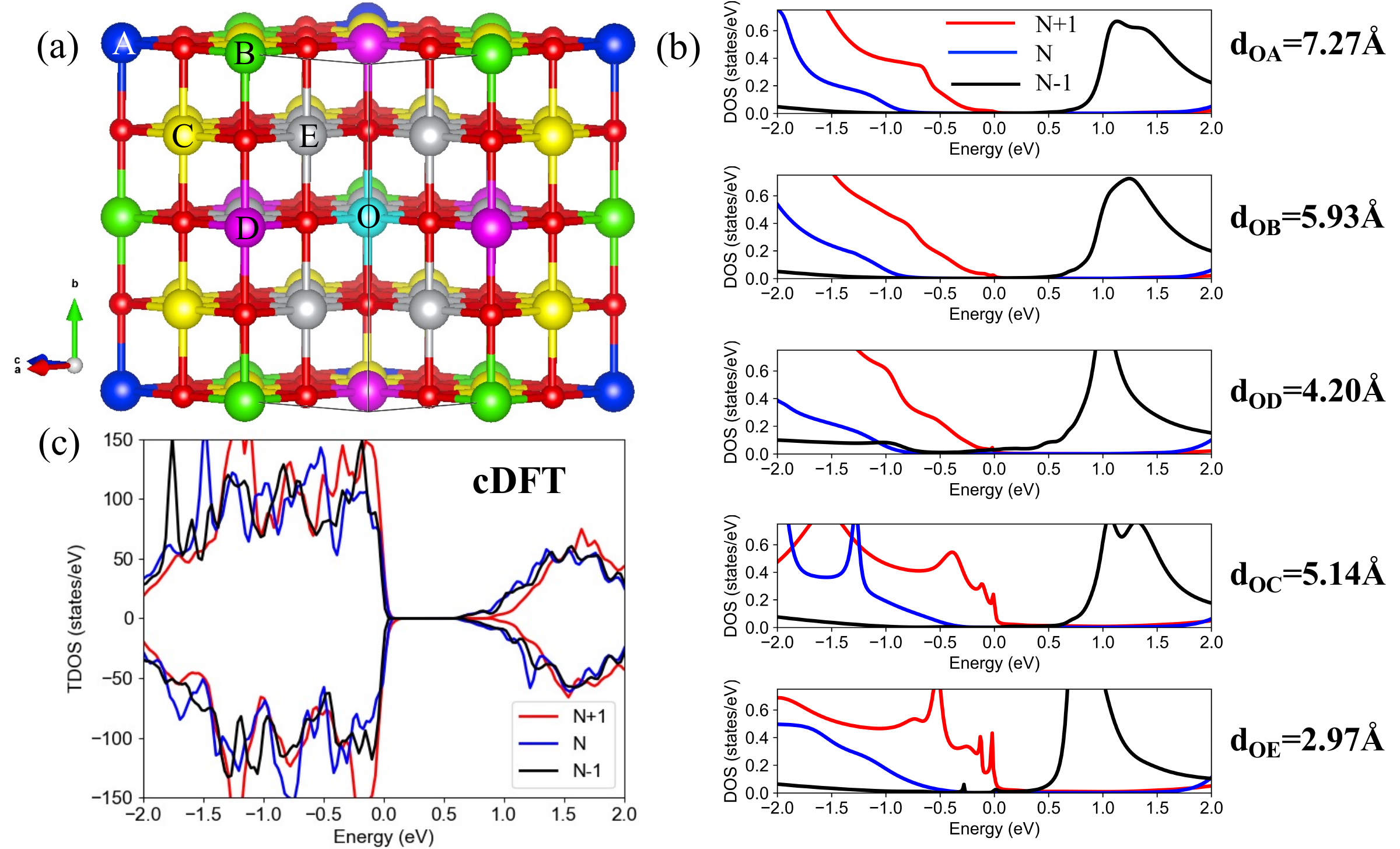}
\caption{(a) Crystal structure of 2$\times$2$\times$2 supercell of NiO. Different colors of balls represent various Ni-ions, whereas the red color denotes oxygen atom. (b) Density of states for various site-dependent Ni-ions obtained by cDMFT calculations for N+1 (in red), N (in blue) and N-1 (in black) configurations, where presence of Kondo-like peak is clearly observed for $d_{OE}$=2.97\AA. (c) Total density of states (TDOS) computed using cDFT for N+1 (in red), N (in blue) and N-1 (in black) configurations. }\label{nio-pdos}
\end{figure*}

\section{Site-dependent PDOS of NiO}

2$\times$2$\times$2 supercell of NiO is shown in Fig. \ref{nio-pdos}(a), where the central Ni ion is denoted by \lq O\rq \,and other five Ni ions are mentioned by A-E. To investigate the effect of Kondo peak as we move away from the central impurity atom, we have plotted the site dependent projected density of states (PDOS) in Fig. \ref{nio-pdos}(b) for different Ni ions located at site A, B, C, D and E. The different distance from the central Ni-atom (`O' site) are also shown in Fig. \ref{nio-pdos}(b). We clearly see the sharp Kondo-like peak is observed for site `E' and `C', which are closer to the central Ni atom due to the strong screening between the conduction electrons and the localized 3$d$ spins of nearby Ni sites. Then the peak fades away as we move from the the impurity atom. As expected, these features are not observed in the cDFT calculation [Fig. \ref{nio-pdos}(c)].

\section{Supercell size dependent $U$}

We have considered elemental Vanadium (V) paramagnetic metal for testing the convergence of $U$ value with respect to various supercell size. The computed values of $U$ using cDMFT method for 2$\times$2$\times$2 and 3$\times$3$\times$3 supercells of V are found to be $\sim$3.8 eV and $\sim$3.6 eV, respectively. It is important to note that a remarkably good matching in theoretical and experimental spectra of V was evident from the previous work of Ref. \cite{sihi2020detailed} using $U$=3.4 eV in eDMFT calculation. We also checked the convergence of the supercell size going from  primitive to $2\times2\times2$ supercell for a represented case (SrIrO$_3$) and found that the $U$ changes only 0.5 eV.

\end{document}